# A NOBLE METHODOLOGY FOR USERS' WORK PROCESS DRIVEN SOFTWARE REQUIREMENTS FOR SMART HANDHELD DEVICES

Tamjid Rahman[1] and M. Rokonuzzaman[2]

[1]Department of Computer Science and Engineering, Stamford University, Bangladesh
[2]Depertment of Electrical Engineering and Computer Science, North South University, Bangladesh

## ABSTRACT

*Requirement engineering is a key ingredient for software development to be effective. Apart from the traditional software requirement which is not much appropriate for new emerging software such as smart handheld device based software. In many perspectives of requirement engineering, traditional and new emerging software are not similar. Whereas requirement engineering of traditional software needs more research, it is obvious that new emerging software needs methodically and in-depth research for improved productivity, quality, risk management and validity. In particular, the result of this paper shows that how effective requirement engineering can improve in project negotiation, project planning, managing feature creep, testing, defect, rework and product quality. This paper also shows a new methodology which is focused on users work process applicable for eliciting the requirement of traditional software and any new type software of smart handheld device such as iPad. As an example, the paper shows how the methodology will be applied as a software requirement of iPad-based software for play-group students.*

## 1. INTRODUCTION

During the last three years, we are witnessing a vast change in the software industry due to customer demand and large revenue of new emerging software such as smart handheld device based software applications, commonly known as Apps. On October 2013, the Verge, an American technology news and media network, reported that Apple has sold 170 million iPads since it launched in April 2010. It is apparent that software Industry is going to be occupied by smart handheld device based software applications. At the same time, it gained attention from software engineering researchers and practitioners.

Requirements engineering (RE) is acknowledged as one of the most important stages in software design and development as it deals with the significant problem of designing the appropriate software for the customer [1]. RE, like all other software engineering actions, must be modified to the needs of the process, the project, the product and the people doing the work [2]. It is therefore obvious that the RE process has important effects for the comprehensive favourable outcome of a software project [4]. Besides this, Software team must scrutinize the perspective of the software work to be carried out, the specific requirements that design and construction must deal with, the precedence that direct the order in which work is to be accomplished [2]. But one might think that RE of traditional software is enough for new emerging software. However, reality does show little evidence. Whether it is traditional or new emerging software development, it is involved with productivity, quality, risk management and validity. Productivity is important for smart handheld device based software because low productivity can limit the range of applicability of the software to any other software.

Software requirements are the key determinants of software quality, certain practical studies shows that errors in requirements are the most numerous in the software life-cycle and also the most costly and prolonged to correct [1]. A study by Boehm and Papaccio [3] exposed that to find and correct an error, it costs US$1 in the requirements definition stage, US$5 in the design phase, US$10in the coding phase, $20US during unit testing, and as much as $200 US after system delivery. RE can reveal developers the source of an error in the requirements definition stage, helping to eradicate possible problems earlier, directing to high quality software. Without quality, customer will not satisfy. As a result, revenue will fall. Furthermore, as usability is a quality aspect found in most arrangements [12][13][14], eliciting usability requirement is likely to be beyond the usability knowledge of most requirements engineers, developers and users. Perry and Wolf have stated that there are static and dynamic limitations on the software components because of usability issues [15]. Even though costs and scheduling can be regarded as productivity concerns, forecasting during the early stages of a project is obviously an issue of risk management [20]. Another important fact is that there are ranges of different opinions on the initial point of RE [2]. To overcome these difficulties, we develop a methodology which is work process driven.

Showing how to overcome the challenges of REP using activities of our methodology, we ensure the validity of the methodology. We expect that the payoff will accurately reflect the customers' needs and thus we will get validation from customers. To show specific example of our methodology, a study is conducted on an English Medium School of Bangladesh.

In section 2, we review the limited study that exists about software requirement of smart handheld device based software and discuss expected payback as a consequence of strict RE practice. Section 3 discusses the important research question. Section 4 discusses the methodology that includes answer of the research question of Section 3. Section 5 describes the RE process, its challenges and payoff of the methodology. Section 6 describes implication for research and practice.

## 2. RELATED WORKS

We looked over the literature of requirement engineering to find requirements engineering process improvement of smart handheld device. There was not much literature about the RE process improvement of smart handheld device. Requirements engineering involves in many ideas, methods and deals with Human Computer Interaction (HCI) especially user-centered design, participatory design and interaction design. However, it is different from HCI in its perspective of the range of design; for example, socio-technical design is hardly ever stated in Requirements Engineering, where the organizational and people part of a system is a clearly defined goal of requirements and design [5]. The main difference is that HCI is mainly concerned how users interact, but before that by RE we decide which contents are appropriate to show. Expert practitioners in HCI are generally designers involved with the realistic application of design methodologies to real-life problems. Researchers in HCI are paying attention in developing new design methodologies, examining with new hardware devices, prototyping new software systems, searching new patterns for interaction and developing models and theories of interaction [17].

The social and cultural aspects of requirements engineering cannot be overlooked which are particularly significant across diverse cultures [18]. For instance, the example we show here is involved with an English Medium School of third world country like Bangladesh. The social and cultural aspects are inevitably different from a first world country like USA. Although a number of issues such as time, result, culture, abstraction and skill may limit use in practice of RE, ethnography has a big deal to present as a technique for RE [19].

However, when technology transfers, in spite of obstacles, there are incentives by means of RE process [6]. But a question may arise that how these incentives and the interaction between the REP and other development processes in figure-1 of [7] are realized as a result of applying our new methodology. The research presented in this paper seeks to answer this question for every incentives and development processes.

## 3. RESEARCH QUESTION

The overall objective of the example discussed in this paper is to innovate iPad based software solutions for improving learning performances of playgroup students. For this purpose, we research on major learning activities of the students and get the following research question:

> *How to develop model of the existing teaching method in the form of activities and their sequence of execution so that we can innovate a software for improving performances of playgroup students?*

To address the above question, we need to find answers of several questions such as

1. What is the role of system elements such as teachers, students, non-computing hardware, computing or electronic hardware, software, database and network in executing each of those teaching activities?
2. What are major key performance indicators (KPIs) and how to measure their current values?
3. How are those KPIs affected by each of the activities?
4. What kind of performance improvement can be anticipated and why?
5. How could it be improved further?

## 4. METHODOLOGY

In fact, to answer the research question of section 3, we have to study the affordance of system elements and this way make an activity model of teaching methods. The term *affordance* refers to the discerned and concrete features of the thing, first and foremost those basic features that specify just how the thing could possibly be used [16].

The answers of the remaining questions can be given as-

Major key performance indicators (KPIs) are given below-
1. When students write any letter, for example A, B, C on their scripts, it looks like

| | |
|---|---|
| A | 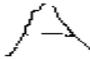 |
| | |

It happens at the beginning stage of writing. We count how much smaller their letter than the actual letter and how often they do that.

2. Usually the hand of students shakes during writing. We count how many times their hands shake.

3. Sometimes students are told to write alphabet A to E. But they write alphabets A, B and then stop writing. The scenario is like-

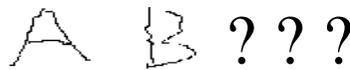 ? ? ?

We also count how many times they do that.

At this stage, we target to improve 100% in the above three situations. To achieve this, we need to design such a software that will improve these performances of students step by step. In order to make progress to meet these objectives, following exercises should be performed.

### 4.1. Observations

According to Yogi Berra, we can observe a lot just by watching. Watching how people do things is a great way to learn their goals and values, and come up with design insight. We divide our observation in two parts. In the first part, we observe all activities which are given in sequence-

- Entering in the class teacher wrote the letter A on the whiteboard and told students that it was A. Then she gave a picture of an apple to the students and told them to write A and paint the apple.
- Students wrote in their books. There was dotted line for number letter A in the books. They just connected all dots. Then they painted the apple.
- Some student wrote the letter correctly, but most students did not because their hands were shaking. Teacher held student's hand and pencil until completion of writing. But every student did not get help. The teacher told them to practice letter A with the help of parents.
- Next, the teacher wrote the letter P on the whiteboard. She followed the same approach like the letter A. This time she gave a picture of pen to all students.
- Then teacher wrote the letter B on the whiteboard. This time she gave a picture of ball to all students.
- Finally, teacher played a video of a cartoon that contained letters A, B and P with apple, ball and pen.
- That time it seemed that students got interested more than the time they were painting the pictures.
- After that, she wrote number 1 on the whiteboard and told students that it was one. She took one stick and one ball and said that there were one stick and one ball in her hands.
- When she taught them number 3, she showed three sticks and ball. We think students learned what is 1, 2 and 3 but they did not understand by seeing sticks and balls because as a play group students they did not know how to count.
- Like alphabets, students practiced numbers using dotted lines on their books.

   - At the last moment of class, teacher gave stickers and chocolates to students who had written well.
   - Some students who did not write well. As a result, they did not get anything.
   - Teacher told them to write well next class, then she will give them sticker and chocolates. After that class was ended.

Now, we move on to the second part in which we describe the observation of successes, breakdown and latent opportunities that occurred when computers were used to support the learning activities. Certain successes occurred, such as students were more motivated by playing videos of cartoon than just showing steel pictures. That is why students did not feel bored (which caused them to come to school and learned more). Some breakdowns occurred, such as it seemed that teacher was ill-equipped to use computers as a teaching tool. The software on computers was not effective. It could be better. A latent opportunity is that a computer network could be used so that teacher can see which students do what from the server. Use of computer software in such a situation could make students more creative. Besides our observation, Yaman claimed the following advantages that will not be acquired by students if computers are not used-

1. Computers help children manage their experience, to set their own pace and to choice the stage of challenge with which they feel at ease.
2. It also helps children to use all of their intelligence to take out facts.
3. Computers attract kids and can capture their full attentiveness, which often causes a deeper attention and concentration. Also allow children to gain knowledge through creating, just as they achieve hands-on knowledge and understanding when they construct forts, make up stories and paint, raise their proficiency.
4. Children build positive attitudes toward technology as they become completely proficient in computers. That will give them benefits over the rest of their lives [25].

### 4.2. Requested Feature of Identified Stakeholders

For change management and feature sizing, we ask all stakeholders to make a list of specific features. However, before that all stakeholders must be identified and different stakeholders have different viewpoints on the significance and priority of requirements and sometimes these views are contradictory. [2][8].

In paper [22], Gause and Weinberg propose the following list of questions to ask the identified stakeholders-
1. Are you the correct person to answer these questions?
  2. Are my questions related to the problem that you have?
  3. Am I asking lots of questions?
  4. Can anyone else provide further information?
  5. Should I be asking you anything else?

These questions will facilitate to "break the ice" and commence the communication that is crucial to successful requirement elicitation.

### 4.3. Discussion with Stakeholders

A question-and-answer meeting layout is not a technique that has been irresistibly effective. As a matter of fact, the Q&A period should be used just for the first meeting and then substituted by a requirements elicitation layout that joins together components of problem solving, negotiation, and specification [2].

An interview with the principle of English Medium School is given in appendix A and in paper [21], a good usability elicitation patterns can be found along with usability mechanism and issues to be discussed with stakeholders.

### 4.4. Find Inspirations (review)

According to the Oxford dictionary, inspiration is the process of being mentally stimulated to do or feel something, especially to do something creative. Five favourite inspirations are given below which are originated from well-known software-

1. We saw an assistant popped up in Microsoft Word when we wrote a letter. In the designed software, there will be an assistant to help the student.
2. Many software including antivirus has the update feature. Time-to-time the antivirus gives users reminder to update. In our designed software there will be auto-reminder options. Suppose, student writes alphabets A and B, but does not write others alphabets. Then the software will remind him/her to write the remaining alphabets.
3. Internet download manager (IDM) shows how much it downloads a file by showing percentage and a bar. Suppose, students are told to write alphabet A to F. They write alphabets A, B and C. Then half of the bar will be filled and besides the bar it shows percentage completed that is 50%.
4. We more or less know the term "parental control". Most parents are busy with their job. They cannot help their children every day. In our designed software, there will be a database which will show how many times a student writes which letters, their performance, improvement and so on.
5. In the class, we observed that if a good student got a sticker, then other students tried to get it. Another fact is that if a student writes well, assistant will give him congratulation. But this type of motivation will not work for a long time. So we have to think another way. If all iPad is connected with a network and a student writes a letter A and B correctly, then other students will see that and try hard to write A and B.

### 4.5. Brainstorm Needs/goals

Using the fact from our careful observation in the class, we go over the findings and use them to brainstorm a list of specific user needs and goals. They are given below—

International Journal of Software Engineering & Applications (IJSEA), Vol.5, No.4, July 2014

1. Instead of existing software, the school needs effective software which will have
   a. Networking features.
   b. Auto-reminder features.
   c. Measurement of improvement.
   d. Feedback features.
   e. Help from assistant.

2. Parents need the parental control feature.

3. Students need to use the software without any help from parents. Overall, they need enjoyment from the software.

4. Teachers need to motivate and encourage the student by involving them in learning to be creative.

### 4.6. Requirement Traceability

To find the origin of requirement and trace how the requirement is changed, requirement documentation is needed. Feature decomposition from requirements documentation is performed so that we can find conformance of the specifications later. These specifications can be consistently relied which is common to most members of the stakeholders as they have a general goal to fulfil.

When managing vast quantities of information, and for complex systems, it is crucial to have some type of automated aid for tracing [23]. Although a good and well-documented example of automatic production of traceability information as a result of development practices is given by Pohl and Jacobs [24], we should develop traceability environments with an accurate mix of automated and non-automated aids, capable of dealing with the functional and non-functional tracing, and with a decent support to all sides of tracing: the definition, the production and capture, and the extraction of traces. The concentration is in solving problems - the problems for which we want to trace information [23].

### 4.7. Information Architecture

As information architecture (IA) is dealt with how people cognitively manage information, IA considerations appear on any product that needs users to make sense of the information offered. Evidently, these considerations are crucial in the case of information-oriented products (like commercial information sites) but they can have an enormous effect even in more functionality-oriented products (like a mobile phone) [26]. When developing IA, we first think about the contents in such a way that groups together them belonging to the same context. Then we have to ensure that the arrangement of the content is useful to meet the user's need. Finally, we ask ourselves is the arrangement aesthetically appropriate interface to attract the users.

The contents of the designed software are given below-
1. All alphabets.
2. Numbers (from 0-9).
3. An assistant.
4. Progress bar with percentage.
5. A picture for each letter and number.
6. Reminder window.

International Journal of Software Engineering & Applications (IJSEA), Vol.5, No.4, July 2014

7. Dotted shape.
8. Music on-off options.
9. Network enabled/disabled options.
10 Pictures of dolls and balloons.

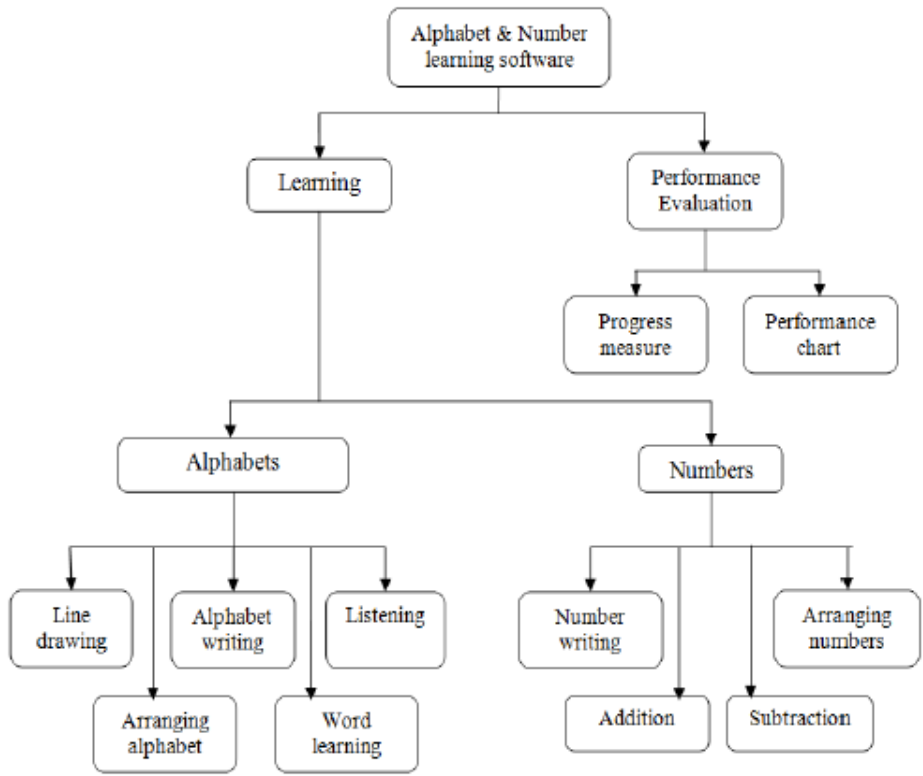

Figure 1. Software content hierarchy.

Although the main user of the software will be playgroup students, teachers and parents can get significant benefits from this software as well in the following way-

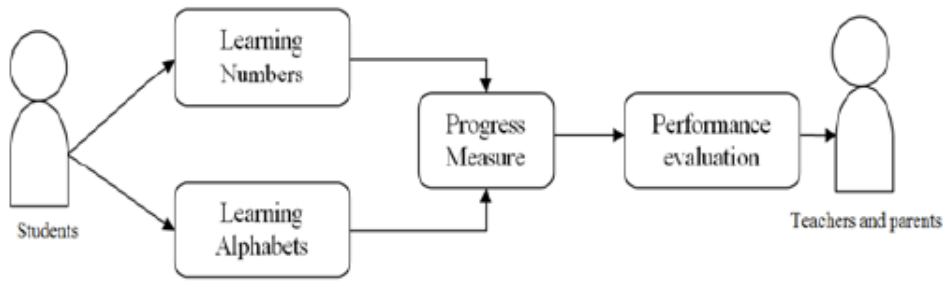

Figure 2. Information about performance passes from students to teachers and parents.

## 4.8 Storyboard

For imagining future or ideal experiences from start to finish, storyboards are a good collaborative tool [27]. Storyboard is built to exemplify design ideas that demonstrate a particular point of view that should distinctly show who the user is, the usage situation, and the user's motivations for using the interface. It should illustrate what the user can achieve with an interface, but it needn't (and often shouldn't) show a particular user interface design [28].

Some storyboards are introduced with some pre-determined elements depending on the stage of the project and the research questions. For example, participants need to add conversations and text explanations on each slot or vice versa, although the designers might have demonstrated some steps already [27]. A good description of storyboard elements and their effects on design process and storyboard consumer experience can be found in paper [29].

In this paper, we introduce a storyboard where we are trying to focus on the child's attention towards iPad and iPad based alphabet learning software. It is important to know how much attractive an iPad is to a child and what the child's view is regarding the interactive software. Here we introduce a scenario in which father buys an iPad for his son to let him play with it as well as use it as learning equipment. His son is amazed after getting an iPad. The colourful alphabets and cartoon picture of the software draw his attention. He finds out that he can draw here by using his finger. During using the software at a stage, he sees an alphabet 'A' which is dotted on the screen and an arrow indicates him to draw the dotted 'A'. He starts to follow the arrow and draw "A" and a colorful apple comes up into the screen, which is very attractive to him as a child. A few moments later he finds a similar thing with a different alphabet 'B' and he already knows what to do. A Storyboard demonstrating the above scenario is attached in the Appendix B.

## 4.9 Wireframe

Wireframing is used to show navigation from one interface to another at the same time showing the content layout to fulfil the user's goal. According to Will Evans, who has over 15 years industry experience in interaction design, information architecture, and user experience strategy, wireframes act as a form of 'thinking device' for the setting and exploration of a given problem space. He also says "designing through the use of wireframes is a search in a problem space of alternatives; it's a process of problem setting as much as it is a process of problem solving which means that I always start with the context".

A good designer must keep in mind the following key points during wireframing:

    1. To help the design team by providing a clear hierarchy of the page content.
    2. To keep our focus on the software structure and user interaction.
    3. To gather new contents what might be needed, but not essential at an early stage.
    4. To express ideas and visualization, which shows the stakeholder what will be the outcome looks like.
    5. To encourage to find out new design and functional features.
    6. To give an overall idea of style, size and formatting to design team.
    7. To be able to find out any missing component.
    8. To minimize any kinds of redundant use.

A wireframe is given in Appendix C.

International Journal of Software Engineering & Applications (IJSEA), Vol.5, No.4, July 2014

## 4.10 Paper Prototype

Paper prototyping can be considered a method of brainstorming, designing, creating, testing, and communicating user interfaces [30]. In [31], steps of paper prototyping and major concern of paper prototyping such as validity, professionalism, development resources can be found. As the methodology is on smart handheld device, the screen size is small. So when preparing the paper prototyping, we have to balance between simplifying functionality and the most common functionality. The success of our prototype not only by how well the interface worked, but also by how much the students learned from it. Few paper prototype screenshots are given below:

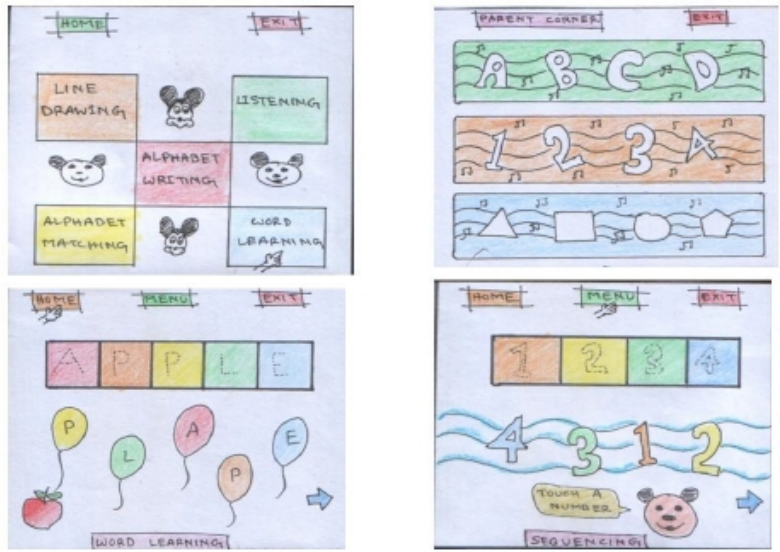

Figure 3. Paper Prototype sample.

## 4.11 Experience Points and User Flow

Great user experiences must have three elements, namely useful, usable and desirable [32]. In [33], although several tactics are given for creating desirable online experiences, they also work for handheld device based software.

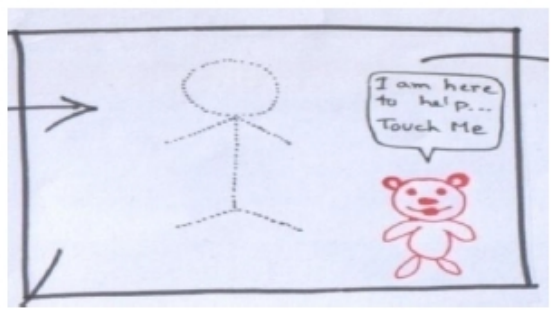

Figure 4. Learning line and circle drawing.

International Journal of Software Engineering & Applications (IJSEA), Vol.5, No.4, July 2014

Here we introduce a way to teach the students how to draw a straight line or circle. We give them to draw some dotted lines and a circle which all together look like a picture of a man. A student might like to draw a picture of man rather than just drawing a straight line.

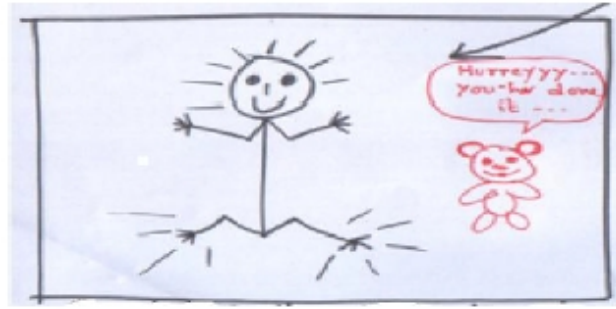

Figure 5. Virtual tutor.

During our observation, we found out that sometimes students need help from teacher or parents. It is not possible to give them help every time they want. So, we introduce a new virtual assistance to help the students during his learning period. An animated character will present on the screen to guide the students. The character will tell the students whether he/she is doing the task correctly or not. It also demonstrates every task if the students seek help from virtual tutor.

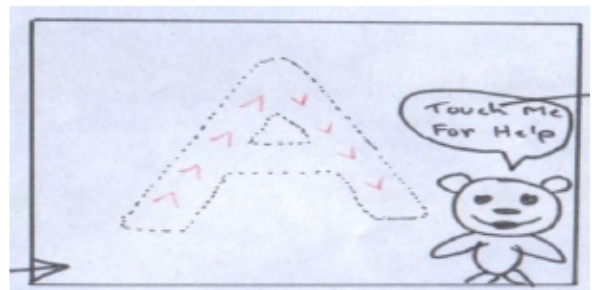

Figure 6: Writing direction.

When a student start writing his first alphabet, it is very normal that he will find it very confusing. So here we introduce a new feature during practice of alphabetic writing. Whenever a student draws a line or circle, another transparent direction will automatically be shown. This feature will help the student to realize which way he should move his finger.

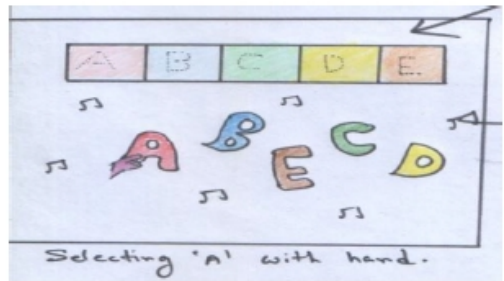

Figure 7: Matching alphabet.

An important experience for the playgroup students is to learn how to arrange them in order. There will be dotted sequenced alphabets at the top of the screen and each alphabet will be in a separate block. Below that, there will be alphabets which are not in sequence and located in a different position on the screen. By touching an alphabet, it will be set in its dotted block. So, students have to maintain the sequence. Touching 'C' after 'A' will not set the 'C' into its block because it is a turn for 'B'.

### 4.12. Perform Heuristic Evaluation

The main goal of heuristic evaluations is to identify any problems (which we call pain point) associated with usability. We want to find that because improvements can be made as part of the iterative design process. During our heuristic evaluation, we found out some of the problem, suggestion and useful feedback. We get some positive feedback on our new innovative ideas like line and circle drawing, virtual assistant, drawing direction etc. We get suggestions which indicate to provide more material, the use of more bold color and so on.

### 4.12.1. Pain Point and Solution

During the observation, we faced some pain points. Those pain points and proposed solutions are given below:

| Pain point | Proposed Solution |
| --- | --- |
| It is hard to draw student's attention towards books. | We remove the boring book and introduce attractive and enjoyable alphabet and number learning software |
| Students get stressed and bored in the classroom | The interface of our design will attract them. They enjoy using the software and thus remove their boredom. |
| Students don't like to draw a straight line or circle at the beginning. | When they complete the drawing using their finger, the virtual tutor will dance and say some encouraging words. By doing this we will be able to keep students likeness. |
| Students get distracted during teaching periods. | Virtual tutor will call them if the user is inactive for a certain period of time. |
| Students need lots of material like, book, notebook, pencil, rubber, sharpener etc. | We provide one stop solution. All you need an iPad and our designed software. Students do not need to worry about pencil, rubber etc. |

## 4.13. Modified Activity Model of Teaching Methods

With the support of our designed software, the following activities and the sequence of execution of those activities can be performed –

1. If students do not pay attention to practice alphabets and numbers, then they will get auto-reminder from the software.
2. Students' parents and teacher can easily know if students do not pay attention for a certain period after getting auto-remainder.
3. When students use the software, they do not need help from anyone except virtual tutor. If Students make a mistake, the assistant will show the mistake.
4. When they perform better, they will get congratulation from the assistant.
5. Students can see the improvement of his classmates simultaneously by the networking features.

Overall, the software will increase involvement of students. We anticipate significant performance improvement because students are encouraged and motivated to write more.

## 5. RE PROCESS, ITS CHALLENGES AND PAYOFF OF THE METHODOLOGY

A number of process phases/activities of software requirement have been proposed in paper [8] [2] [9] [10]. Although there is no universal requirement process and informal processes remain in nearly all software industry, the most common are requirements elicitation, requirements analysis, requirements negotiation, requirements documentation and requirements management.

Two more requirement processes are found, namely specification and validation.

In requirements elicitation activity, software engineers work with customers and system end-users to discover the application area, what services the system should offer, the necessary performance of the system, hardware constraints, and so on [8].

The requirements analysis phase is involved with examining the set of elicited requirements for conflicts, ambiguities, overlaps, omissions and inconsistencies [9].

It is usual for customers and users to demand for more than can be obtained, given insufficient business resources. They also propose conflicting requirements, arguing that their version is crucial to their special needs. Using a process of negotiation, the requirements engineer must resolve these conflicts [2].

The documentation of requirements could be presented as the activity of creating and maintaining the documents associated with the process [9] and when a document is used as a communication medium between the customer and supplier, it is called requirement specification [35]. Characteristics of a good requirement specification are correct, unambiguous, complete, consistent, ranked for importance and/or stability, verifiable, modifiable and traceable [34].

The process of understanding and controlling changes to system requirements is requirements management and the process of verifying that requirements essentially define the system that the customer actually wants is requirements validation [8].

Now the following table presents the challenges in REP and activities to overcome them. It also presents expected payoff which can be achieved by applying those activities of the methodology-

| REP | Challenges in REP | Activities to overcome challenges | Expected payoff |
| --- | --- | --- | --- |
| Requirement elicitation | System boundary is ill-defined [11]. | Making activity model of teaching methods. | Accurate estimates |
| | Given unnecessary design information by users [11]. | Observation and discussion with stakeholder. | |
| | Users have incomplete understanding of their needs [11]. | Brainstorming needs. | Improved feature coverage. |
| Requirements analysis | The lack of appropriate application domain knowledge and the context-situated character of information make the designers unable to think in the same terms as the user [9]. | Information architecture. | Early error detection and reduced rework. |
| Requirements negotiation | Stakeholders use different terminology in expressing requirements and this can create an apparent conflict [9]. | Requested feature of identified stakeholders and discussion with stakeholders | Effective project negotiation |
| Requirements documentation | Dilemma about which level of formality contained in Requirements documentation [9]. | Requirement traceability | Improved communication, solid foundation for later phases |
| Requirement management | To keep track of individual requirements and maintain links between dependent requirements so that impact of requirements changes can be assessed [8]. | Requirement traceability | Reduced requirements creep, improved Management |
| Requirements specification | Ease of omitting "obvious" information [11]. | Build storyboard, wireframe model, paper prototype | Accurate estimates |
| Requirements validation | Users need to imagine how that system would fit into their work [8]. | Build storyboard, wireframe model, paper prototype, | Fewer defects |
| | Errors in a requirements document are discovered during development or after the system is in service [8]. | Experience points and user flow, Perform Heuristic Evaluation | Reduced rework |

## 6. IMPLICATIONS FOR RESEARCH AND PRACTICE

This research has several important implications for both researchers and practitioners. First, we believe that our methodology has a significant role to fulfil the gap in traditional software requirement engineering and appropriate for the software of smart handheld device such as iPad. Investigating the challenges of REP and the activities to overcome them and expected payoff to gain productivity, quality and risk management as the literature suggests has turned out to be a complicated but worthwhile task. Using the fact from observation of the methodology, both requested feature of identified stakeholders and discussion with stakeholders show how requirements elicitation can be performed that have had an impact throughout the methodology. Though discussion with stakeholders and of the methodology introduced collaboration with stakeholders, observations and requested feature of identified stakeholders of the methodology will be different for diverse culture. So these parts of the methodology can be changed. Another direction for research is that how to develop instruments for more objectively assessing and measuring steps of the methodology. For example, when we perform heuristic evaluation, the evaluation can be performed by several peer-reviews who will be asked to evaluate the proposed design of the software against a set of predefined criteria so that later we can compare their judgments to reach optimum evolution. During peer reviews, the research question is what the relevant and meaningful criteria are to judge improvements. Another research question is that is it possible to apply this methodology as a solution to other new innovative software? These findings produce interesting new directions for research.

## 7. CONCLUSIONS

It is our belief that the software based on our methodology will facilitate communication between teacher and students performing different types of alphabet learning games, attractive educational animation to attract the children, class test, practice material etc. The software also contains a database containing the day to day activities, student's information, course description and materials. More specifically the software will help teachers to provide more interactive learning for the students. If students practice with our designed software and parents and teachers supervise their performance at least weekly, performance can be improved further.

We expect that when a complete software is developed using our methodology, it will dominate in the app market. As a result, substantial revenue will be earned.

## Appendix A

During our interview session we asked following questions:

Q: What do teachers in this school teach playgroup student at the very beginning?
A: At first, all students recite various rhymes with teachers. Music is played at that time. This situation continues for 2-3 classes so that students can memorize the rhymes and recite them solely.
Q: After the rhymes, teachers teach the students how to write alphabets, right?
A: Actually, they don't. Teachers make them familiar with all alphabets.
Q: How do they do that?
A: Teachers show various pictures, sticker and videos to students with appropriate letter. For example, teacher shows a picture of ball for letter B. We follow the same method for numbers.
Q: In class we observed that the teacher taught the letter P before B. Why was that?
A: It was teaching strategy for the convenience of students. If they learn how to write letter P, then it will be easy for them to write letter B.
Q: When they are familiar with alphabets what is the next step?

International Journal of Software Engineering & Applications (IJSEA), Vol.5, No.4, July 2014

A: After making sure that student can easily recognize alphabets, teachers help them to write the alphabets. For convenience, capital letters are taught at first. Some students do not need help but the letter they write is not much good. We help them as well.
Q: How do teachers help weak students to write alphabets?
A: When a weak student writes a letter, his hand usually shakes. That time teacher holds his hand and pencil until completion of writing. Several times teacher helps the student in this way.
Q: How do teachers encourage students when they write alphabet well?
A: For encouragement, teachers give them sticker as a reward. Sometimes teachers told the class to give a big clap for the students.
Q: Suppose, a student writes a letter, but not much good. Another student writes it very well. How do teachers reward them?
A: If a student writes a letter well, teachers give him a big sticker and tell him thank you very much. If a student writes a letter that does not look quite good, teachers give him a small sticker, show the right way to write the letter and then tell him to write it.
Q: Do teachers face any problem when she teaches them to write?
A: Some students press the pencil on the script more than expected. That time lead of the pencil breaks and student gives the pencil to the teacher to sharp. Teacher sharps it and gives the student. But student breaks lead of the pencil again. This problem happens again and again.
Q: Which strategy is important in your point of view to improve students' performance?
A: Personally, I emphasize on involvement. The more students involve in writing, the more improvement we will observe. I mean we always encourage them to involve in writing.
Q: As we are going to design iPad based innovative software solutions for play group students, what do you expect from us?
A: We expect software that will help students when there are no teachers or parents. The ultimate goal of software will make students more creative.

## Appendix B

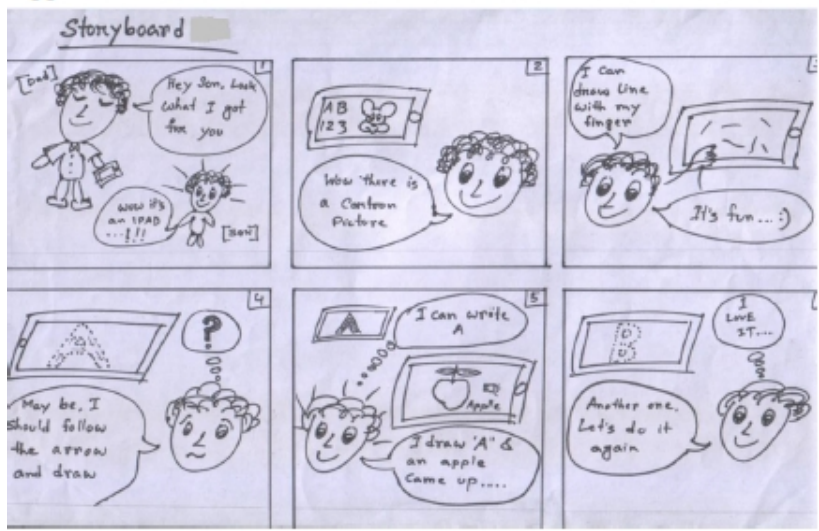

36

## Appendix C

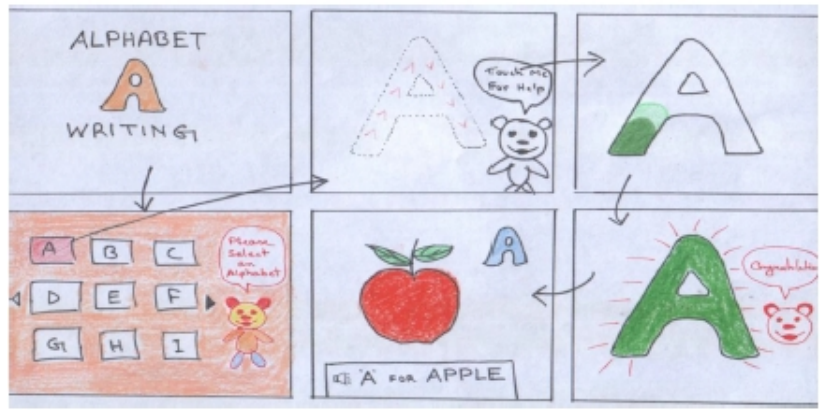

**Authors**

**Tamjid Rahman** is a faculty of Stamford University, Bangladesh. He received his BS degree in Computer Science and Engineering from Stamford University, Bangladesh in 2010. He is a MS student of North South University. His research interest lies in Artificial Intelligence, Software Engineering, Programming 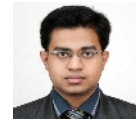

**Dr. M Rokonuzzaman** is a faculty of North South University. He received his Ph.D. in EEE from Memorial University of Newfoundland, Canada. His research interest lies in Software Engineering, Robotics, Computer Vision.

.